\documentclass[apj]{emulateapj}

\usepackage{graphics,graphicx,rotating}
\usepackage{natbib}
\usepackage{xspace}
\usepackage{amssymb}
\usepackage{amsmath}
\usepackage{epstopdf}
\usepackage{subfigure}
\usepackage{epsfig}
\usepackage{color}

\newcommand{\Msun}{ {\ensuremath{\mbox{M}_{\odot}}} }

\newcommand{\ECvir}{ {\ensuremath{\mbox{C}^{e}_{vir}}} }

\newcommand{\EMvir}{ {\ensuremath{\mbox{M}^{e}_{vir}}} }
\newcommand{\Rvir}{ {\ensuremath{\mbox{R}_{vir}}} }
\newcommand{\ERvir}{ {\ensuremath{\mbox{R}^{e}_{vir}}} }
\newcommand{\qDM}{ {\ensuremath{q}} }
\newcommand{\Peq}{ {\ensuremath{P_{eq}}} }
\newcommand{\Pth}{ {\ensuremath{P_{th}}} }
\newcommand{\Pnth}{ {\ensuremath{P_{nth}}} }
\newcommand{\Rpr}{ {\ensuremath{R_{pr}}} }
\newcommand{\Rnpr}{ {\ensuremath{R_{npr}}} }

\begin{document}

\title{Testing Hydrostatic Equilibrium in Galaxy Cluster MS 2137}

\author{
I-Non Tim Chiu\altaffilmark{1} and Sandor M. Molnar\altaffilmark{1}
}

\altaffiltext{1}{
Leung Center for Cosmology and Particle Astrophysics, 
National Taiwan University, Taipei 10617, Taiwan, R.O.C.
}

\begin{abstract}

  We test the assumption of strict hydrostatic equilibrium in galaxy cluster MS2137.3-2353 (MS~2137) using the latest CHANDRA X-ray observations and results from a combined strong and weak lensing analysis based on optical observations. 
  We deproject the two-dimensional X-ray surface brightness and mass surface density maps assuming spherical and spheroidal dark matter distributions.
  We find a significant, 40\%--50\%, contribution from non-thermal pressure in the core assuming a spherical model.
  This non-thermal pressure support is similar to what was found by \cite{Molnar2010} using a sample of massive relaxed clusters drawn from high resolution cosmological simulations.  
  We have studied hydrostatic equilibrium in MS~2137 under the assumption of elliptical cluster geometry adopting prolate models for the dark matter density distribution with different axis ratios. 
  Our results suggest that the main effect of ellipticity (compared to spherical models) is to decrease the non-thermal pressure support required for equilibrium at all radii without changing the distribution qualitatively. 
  We find that a prolate model with an axis ratio of 1.25 (axis in the line of sight over perpendicular to it) provides a physically acceptable model implying that MS~2137 is close to hydrostatic equilibrium at about 0.04--0.15 \Rvir and have an about 25\% contribution from non-thermal pressure at the center.
  Our results provide further evidence that there is a significant contribution from non-thermal pressure in the core region of even relaxed clusters, i.e., the assumption of hydrostatic equilibrium is not valid in this region, independently of the assumed shape of the cluster.

\end{abstract}


%
%

\section{Introduction}

  Galaxy clusters are the largest gravitationally bound and virialized systems in our universe. 
  They are important cosmological tools to probe the curvature, mass and energy density of the universe. 
  One of the most important parameters of galaxy clusters for cosmology is the total mass. 
  Cosmological simulations make specific predictions for the total mass distribution of clusters and the distribution of mass in individual clusters. 
  The total mass of clusters is a particularly important scaling parameter when using galaxy clusters to constrain cosmological parameters. 
  Unfortunately, the mass of clusters can not be observed directly, it has to be derived from observations with some additional assumptions.
  The most widely used (and independent) methods to determine cluster masses are based on gravitational lensing and X-ray observations. 
  The method based on gravitational lensing is independent of dynamics because the masses is derived from the tangential shear of the background galaxies distorted by the gravitational potential of the cluster. 
  On the other hand, cluster masses estimated from X-ray observations is based on dynamics.
  The X-ray method assumes that the gas is relaxed and in strict hydrostatic equilibrium (i.e. the cluster is supported only by the thermal pressure). 
  However, there is a long-standing problem that these two methods sometimes result in significantly different cluster masses \citep{Wu1998,Allen1998,Hicks2000,Hoekstra2007,Piffaretti2008,Peng2009,Riemer2009,Meneghetti2010, Zhang2010,Okabe2010,Morandi2011,Morandi_Pedersen2011}.

  The most popular method to derive the total mass of relaxed clusters from X-ray observations assumes that they are in strict hydrostatic equilibrium and have spherical symmetry. 
  Unfortunately, there are several physical processes which would break the strict hydrostatic equilibrium in clusters, for example: turbulent gas motion, cosmic rays, heating, cooling, and magnetic fields \citep{Ensslin1997,Norman_Bryan1998,Rasia2004,Sarazin2004,Rasia2006,Nagai2007,Pfrommer2007_1,Pfrommer2008_2,Pfrommer2008_3,Guo2008,Skillman2008,Fang2009,Lau_Kravtsov_Nagai2009,Vazza2009,Lagana2010,Valdarnini2011,Parrish2012}.

  Numerical simulations suggest that turbulent motions of the intracluster gas (ICG) caused by mergers and shocks provide a significant non-thermal pressure support which varies as a function of radius \citep{Norman_Bryan1998,Dolag2005,Vazza2006,Rasia2006,Iapichino2008,Lau_Kravtsov_Nagai2009,Vazza2009,Molnar2010}. 
  Based on smooth particle hydrodynamics (SPH) simulations, \cite{Vazza2006} showed that the turbulence energy of gas particles scales approximately with thermal energy of clusters. 
  In the outskirts of clusters, the ICG is also supported by the bulk gas motion and turbulence created by accretion shocks \citep{Burns2010}.
   The scaling relations between the thermal and turbulence energy in the different radii among the different redshifts have been investigated by \cite{Vazza2011}. 
   Their results suggest that the non-thermal energy increases with radius and it may be more than 20\% of the thermal energy near the \Rvir.
   \cite{Parrish2012} conducted magnetohydrodynamics simulations to study the turbulences deduced from the magnetothermal instability in the outskirts of the clusters, and suggested that non-thermal pressure can provide 5 -- 30\% support against gravity beyond $R_{500}$.
  \cite{Lau_Kravtsov_Nagai2009} studied the effect of gas motion in the mass estimation of simulated clusters, and concluded that the non-thermal pressure due to residual gas motion at large radii, if not taken into account, would result an underestimate of the cluster mass of about 15\% confirming previous results.
  Based on an analysis of massive clusters drawn from cosmological simulations, \cite{Molnar2010} suggested that the non-thermal pressure can provide a significant, about 30\%, support in the central region, a minimum support at about 0.1--0.2 \Rvir, and increases with radius to about 35\% at \Rvir.
  \cite{Fang2009} suggest that the pressure from rotation and streaming motions are significantly comparable to the pressure raised from random turbulent pressure out to certain radius range. 
  The hydrostatic equilibrium assumption could result an underestimate of the mass of clusters due to the neglected contribution from turbulent bulk motion \citep{Rasia2006}.
  The mass estimation of the cluster could be fairly reconstructed if the gas motion pressure is taken into account in the hydrostatic equilibrium equation \citep{Rasia2004,Lau_Kravtsov_Nagai2009}. 
  In addition, inappropriate assumption of the gas and the dark matter halo distributions could also bias the mass estimation of the clusters \citep{Gavazzi2005,Corless2008,Oguri2010,Morandi2011,Morandi_Pedersen2011,Morandi_Limousin_Sayers_2011,Sereno_Umetsu2011,Sereno2012_a,Sereno2012_b}.

  On the other hand, gravitational lensing of background galaxies is an unique, direct probe of the distribution of matter in galaxy clusters, determining the mass profiles from two-dimensional (2D) projected physical quantities such as reduced shear for example. 
  However, it is extremely hard to fully reconstruct the three-dimensional (3D) mass distribution even with triaxial models fitting to the lensing data, because the lensing technique is sensitive to all the projected mass in the line of sight (LoS) \citep{Corless2008}.
  Methods based on lensing assuming spherical symmetry would possibly overestimate (underestimate) the cluster masses if the mass distributions are prolate (oblate) with the non-degenerated principle axis aligned in LoS (see for example \citealt{Gavazzi2005}). 
  Moreover, in order to derive accurate mass profiles at all radii, it is necessary to combine strong and weak lensing observations which are sensitive to small and large radii.

  Molnar et al. (2010, hereafter M10) studied Abell~1689 using the latest CHANDRA data and three massive clusters of galaxies drawn from cosmological adaptive mesh refinement (AMR) simulations. 
  Assuming spherical geometry, they found a significant, about 45\% and 35\%, contribution from non-thermal pressure in the core regions of Abell~1689 and simulated clusters respectively.
  \cite{Morandi2011} fitted triaxial dark matter mass models with their major axis aligned with the LoS to the full lensing and X-ray data simultaneously, and derived the dark matter axis ratio is 2.02 $\pm$ 0.01 (1.24 $\pm$ 0.13) along the LoS (in the plane of sky) for Abell~1689. 
  But, unlike M10, \cite{Morandi2011} assumed that the non-thermal pressure contribution is constant throughout the all radii. 
  They found that the non-thermal pressure contributes about 20\% to the total pressure needed if the hydrostatic equilibrium holds. 
  Using similar methods, \cite{Morandi2011} and \cite{Morandi_Limousin_Sayers_2011} presented a triaxial model analysis fitting simultaneously to the multiple-wavelength data of Abell~383, and Abell~1835, and derived the non-thermal pressure contribution, about 11\% -- 20\%, out to $R_{200}$.

  In this paper we continue our previous analysis of non-thermal pressure support in relaxed clusters assuming more general, non-spherical (ellipsoidal) geometry for the matter distribution.
  Since we are interested in the physics of the mass discrepancy, instead of comparing the total masses of clusters determined from X-ray and lensing observations, we test (as in M10) the assumption of strict hydrostatic equilibrium by comparing the two competing physical processes, self-gravity, acting inward, and the gradient of the pressure, acting outward. 
  In particular, we study strict hydrostatic equilibrium in cluster MS~2137.

  MS~2137 ($z=0.313$) has been studied extensively by using X-ray and optical (lensing) data \citep{Hammer1997,Gavazzi2005,Comerford2006,Shu2008,Sand2008,Merten2009,Donnarumma2009,Comerford2010}. 
  \cite{Comerford2006} modeled the central mass distribution of MS~2137 assuming an asymmetric NFW profile with the observed gravitational arcs. 
  They found that the reconstructed lens system and the estimated mass profile are in a good agreement with simulations. 
  \cite{Donnarumma2009} (hereafter D9) estimated the mass of MS~2137 in the central and the outer regions by analyzing strong lensing data observed by Hubble Space Telescope and CHANDRA X-ray data, respectively. 
  Assuming spherical and elliptical cluster models with the major axis aligned in the LoS, \citealt{Gavazzi2005} (hereafter G5) derived the total mass profile in MS~2137 between 0.02 and 1.0 Mpc by fitting the conventional and general NFW models to the strong and weak lensing data. 
  G5 found a better agreement between masses determined from X-ray and lensing observations using prolate models.
  G5 carried out a detailed combined weak and strong gravitational lensing analysis of MS~2137 (the same data were used by D9), therefore we use their results as our reference for lensing.

  MS~2137 is a relaxed cluster with a round morphology (close to circular) in projection (in the plane of the sky). 
  In order to simplify the modeling, in this case, with a good approximation, we can derive the thermal and equilibrium pressure profiles for this cluster assuming spherical or the elliptical (prolate) models with the non-degenerated axis aligned in the LoS.

  The outline of our paper is as follows: 
  in section 2, we discuss our X-ray data analysis of CHANDRA data for MS~2137, and the derivation of the thermal pressure, \Pth, based on our analysis.
  We discuss our method to derive the equilibrium pressure, \Peq, in section 3, based on published results from strong and weak gravitational lensing. 
  We present our test of the assumption of hydrostatic equilibrium in MS~2137, we discuss the possible sources of non-thermal pressure support and systematic biases in section 4 and draw our conclusion in section 5. 
  Throughout in this paper, we assume a concordance cosmological model with $\Omega_m$ = 0.3, $\Omega_\Lambda$ = 0.7, and $H_0$ = 70 km s$^{-1}$ Mpc$^{-1}$. 
  Based on this model, the angular size 1$\arcmin$ = 0.275 Mpc for MS~2137.
  Unless otherwise stated, all errors, error bars and dashed lines are represented 1$\sigma$ confident intervals.

%
%
\section{X-ray data analysis}

\subsection{X-ray Data Reduction and Cleaning}

  We use publicly available observations of MS~2137 from the CHANDRA Data Archive (CDA). 
  We found three observations with observation IDs (ObsID) 928, 4974 and 5250 with exposure times of 44.17 ks, 58.14 ks and 41.09 ks, respectively. 
  We reduced these three data sets by using the latest CHANDRA data analysis software package CIAO-4.4 and updated complement calibration CALDB-4.4. 
  We followed the standard ACIS data preparation\footnote{http://cxc.harvard.edu/ciao/guides/acis\_data.html} to reprocess all observations from level 1 event files (evt1) to level 2 reduced event files (evt2).
  First, in order to improve the quality of our data, we removed the streak by running the script \textit{destreak} before we created the new bad pixel files. 
  Then, we created the new bad pixel files correcting the known bad pixels and columns, identifying and marking the afterglow and hot pixels by running the scripts \textit{acis\_find\_afterglow} and \textit{acis\_build\_badpix}. 
  Our next step was to run the script \textit{acis\_process\_events} producing a new evt1 with event grades 0,2,3,4 and 6 using the latest calibration data including the new bad pixel file created previously. 
  Then we applied two stages of filtering to these new evt1 files: first: for the imaging observation, we filtered for bad grades in the imaging observation, second: we used the primary Good Time Interval (GTI) to create the new evt2 file for each observation data sets. 
  In addition, we detected and removed the point sources by the script \textit{celldetect}. 
  Only a small number of the contaminated point sources was found resulting the removal of only a few pixels.

  Most of the contamination can be filtered out after the data reprocessing, but contamination from strong background flares, which could increase the count rate several times, still has to be considered.
  Our final step was to remove periods of high count rates due to background flares based on the light curves of each observation.
  However, the three observations have a different amount of contamination from strong flares.
  D9 studied MS~2137 by the joint-analysis of the arcs due to lensing observed by HST and CHANDRA X-ray data, but they only used ObsIDs 928 and 5250, stating that there is a difficulty in identifying the time intervals contaminated by background flares in the observation with ObsID 4974.
  Furthermore, the net exposure times in their work were reduced from 44.17, 58.14 and 41.09 ks to 20.07, $\sim$ 16 and 25.0 ks for ObsID 928, 4974 and 5250, respectively. 
  As a result, they decided to discard ObsID 4974 in order to avoid the possibility of a systematic bias due to contamination from strong background flares.
  However, because the net exposure time contributed by ObsID 4974 is still considerable to the total net exposure time after cleaning, including ObsID 4974 can decrease the statistical uncertainty in our results.
  Thus, after a careful flare-cleaning process, we decided to include ObsID 4974 in our study.

  The cleaning procedure for strong flare background for all three observations we adopted was the following.
  After the reprocessing the data creating evt2 files from the evt1 files, the light curves of ObsID 5250 show no difficulty in filtering the background flares via running the CIAO-4.4 software script \textit{lc\_clean}.
  ObsID 928, on the other hand, is clearly contaminated by strong flares during the observation time about 59332 to 59352 ks, while the rest of exposure time shows a rather stable count rate.
  Therefore we discarded the time interval between 59332 and 59352 ks for ObsID 928 before we run the script \textit{lc\_clean}.
  Parts of the S3 chip, 5$\arcmin$ away from the emission peak, are selected as background fields in the script \textit{lc\_clean} for ObsIDs 928 and 5250, and the reduced net exposure times are consistent to the result of D9.
  In ObsID 4974 (with the net exposure time $\sim$ 60 ks), we find strong contamination from background flares, which is also described in D9. 
  In order to probe contamination due to flares in ObsID 4974, we extracted the light curves from 5 annuli centered in the emission peak with radii uniformly distributed between 0.5$\arcmin$ and 4.5$\arcmin$ in the log space. 
  We find that the light curves in these five regions show a similar trend: the count rate of the first $\sim$ 40 ks observation time is anomaly higher than adjacent exposure time after that, and the count rate of the later exposure times, about 20 ks, is more steady implying that this interval suffers from fewer flares. 
  Based on our findings, we concluded that the background scattering distributes approximately uniformly near the field of the cluster in the S3 chips but not regularly and periodically spreading during the whole observation time. 
  Other words, the background flares highly bias the cluster emission in the first $\sim$ 40 ks of the exposure, but much less after that.
  Hence, we decide to analyze only the last 20 ks of the observation time for ObsID 4974.
  We further filtered the flares in this time interval using the region of the annulus, having the inner and outer radii of 2.89$\arcmin$ and 4.5$\arcmin$, respectively, centered on the emission peak as our background criteria via running the script \textit{lc\_clean}. 
  The GTIs for these three observations were determined after all corrections had been applied.
  The net exposure times after applying these GTIs are reduced from 44.17, 58.14 and 41.09 ks to 14.07, 16.633 and 15.353 ks for ObsIDs 928, 4974 and 5250, respectively.

\bigskip

\subsection{Background Subtraction}

  The usual method to separate the X-ray emission due to galaxy clusters from the background is to subtract the background as determined from blank sky observations from the data.
  However, CHANDRA observations have two telemetry modes, FAINT and Very FAINT (VFAINT) modes, which have different grading selections. 
  The standard blank sky background provided by the CHANDRA science team are reproduced from observations processed using the FAINT modes, while all three observations we are considering were using the VFAINT mode. 
  Therefore, instead of making use of the blank sky backgrounds, we used local backgrounds with the assumption that the quiescent backgrounds did not vary significantly between the CCD chips. 
  For each observations with ObsIDs 4974 and 5250, the background was extracted individually from another back-illuminated chip (S1), which covers a part of the sky far from the cluster center, thus it can be considered free of cluster emission. 
  Because of the lack of the S1 chip in ObsID 928, we extracted the local background from S3 far away from the cluster center.
  As a consistency check, we compared this local background of ObsID 928 to the background extracted from S1 chip in ObsID 5250, which the data set suffers from less background flares. 
  We found spectra of these two backgrounds are consistent with each other after normalization of the fluxes.
  Thus we apply this to background selection for ObsID 928 and the local background of ObsID 928 is selected as the region of residual S3 chip having the distance of 4.5$\arcmin$ far from emission peak.

%
%
\begin{figure}
\centering
\includegraphics[angle=-90,width=0.485\textwidth]{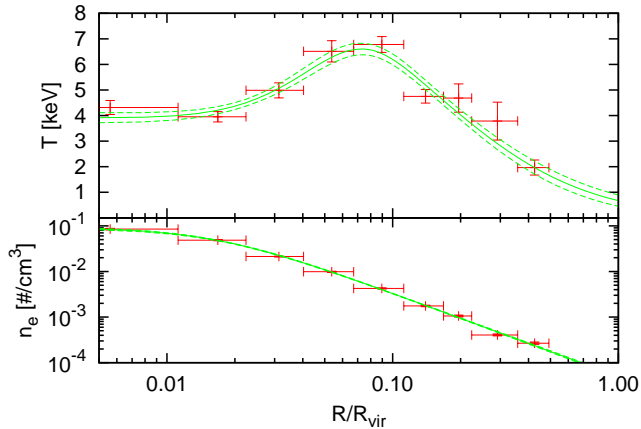}
\caption{
  The deprojected temperature (upper panel) and the electron number density profile (lower panel) assuming spherical symmetry.
  The red color dots are the results of the simultaneously fitting combining three observations and assuming spherical symmetry.
  The green curves are the best-fitted functions.
\label{TNe}
}
\end{figure}

\bigskip

\subsection{Spectral Fitting}

  We extracted 9 circular annuli (8 complete and 1 fragment of the annulus having the spanning angle of 265$^{\circ}$ due to the lack of the field) centered in the X-ray emission peak $(21^{h}40^{m}15.2^{s},-23^{\circ}39\arcmin 40\arcsec)$ for all three data sets. 
  We use the CIAO-4.4 software tool \textit{specextract} to create the spectrum, the Response Matrix File (RMF) and the Ancillary Response File (ARF) for each region. 
  Since MS~2137 is located at the center in one of the two back-illuminated (BI) chips, S3, in all three ObsIDs, we had no difficulty in extracting the spectra, RMFs and ARFs out to the radii of 220$\arcsec$.
  The background spectra were extracted using the CIAO-4.4 software tool \textit{dmextract} independently with the source spectra.

  We deproject the observed 2D spectral image to get the 3D gas properties.  
  We adopt two types of models for the distribution for the ICG: spherical and elliptical. 
  Assuming circular symmetry in the plane of the sky and using a prolate model with the non-degenerated axis aligning in the LoS, we divide the gas distribution into 9 shells according to the 9 annuli we used in the spectral extraction. 
  Again, we assume the gas is homogeneous in each shell.

  We use the X-ray fitting package \textsf{XSPEC-12.5} in our spectral analysis. 
  We describe the gas emission in each shell by an optically thin plasma emission model, \textit{MEKAL}, multiplied by the photo-electric absorption model, \textit{WABS}.
  Our model for the projected spectra is generated by using the appropriate volume ratios associating the normalization parameters in all \textit{MEKAL} models (see Appendix for more details of this projection).
  We fixed the galactic photoelectric absorption\footnote{column density investigated from http://heasarc.gsfc.nasa. gov/docs/tools.html} and the redshift to the value 3.55 $\times 10^{20} \mbox{cm}^{-2}$ and 0.313, respectively. 
  We fitted for the three dimensional distribution of the temperature, abundance and normalization parameters using C-statistic in the energy band 0.5 to 7.0 keV.

  As a consistency test, first we fitted the spectra for each observation individually to see whether the results agree within errors. 
  The temperature profiles for ObsID 4974 and 5250 are consistent with the result of D9 except that our maximum values are slightly higher than theirs (we will discuss it in the next paragraph).
  On the other hand, the obtained deprojected temperatures and densities did not agree in the outer regions (above distance 1.3$\arcmin$ far from the emission peak) for ObsID 928, although the inner regions were consistent with the results derived from the other two ObsIDs and D9.
  We performed fittings to spectra in the outer regions (farther than 1.3$\arcmin$) tentatively only in the 1.0 to 7.0 keV band (where the energy bands of background spectra are the most consistent between the local and the blank sky backgrounds) using the blank sky background scaled to ObsID 928. 
  However, we found that there is no significant differences in the spectral fitting results between using the scaled blank sky background and the local background we selected (extracted from the region beyond 4.5$\arcmin$) and the fitted energy ranges. 
  The spectral fitting remained extremely ill-constrained in the outer regions of ObsID 928 possibly because the emission is dominated by the background.
  In order to avoid bias in our final results, we decided to discard these regions (about 1.3$\arcmin$ away from emission peak) of ObsID 928 when we performed simultaneous fitting to all observations (i.e. we discarded the last three annuli in ObsID 928).

  We also compared the results of our X-ray data analysis to those of D9.
  There is a good agreement between the two density distributions (within errors). 
  We compared our spectroscopic-like temperature profile projected from our best-fit 3D model using the weighting derived by \cite{Mazzotta2004} to the deprojected temperature profile derived directly from observations by D9 (see red points in their Figure 3).
  Our temperature profile is also consistent with that of D9 except at about 200 kpc where our temperature is 1 keV higher than theirs.
  This discrepancy is most likely coming from two sources:
  i) We sample the outer regions (100--700 kpc) with more radial bins than D9 (we are using 4, D9 are using 2 bins). 
  At around 200 kpc there is a large temperature gradient which, if averaged over a large radial bin, results in a lower ring averaged temperature. 
  ii) Statistical fluctuations.

  We found that there is no significant change in the distribution of the temperature and density in our X-ray analysis for the assumed axis ratios for the dark matter mass distribution, thus we only present our deprojected temperatures and electron number densities for the spherical case (See Fig.~\ref{TNe}).
  We use the virial radius, \Rvir, as a distance unit in our figures to make it easier to compare our results to those of others (see the Appendix for the exact definition of our distance unit).

\bigskip

\subsection{Derivation of Thermal Pressure}

  We use the following smooth parametrized functions for the deprojected temperature and density distributions and perform fittings to the data points we obtained as described in the previous previous section: 
\begin{eqnarray}
\label{smthT}
T_{3D}(m_g(R,Z))  &=  &T_0\frac{ 1+g_c \, \exp \{-(m_g/r_c)^{a_c} \} } { (1+m_g/r_T)^{a_T} } \\
\label{bmodel}
n_e(m_g(R,Z))     &=  &\frac{n_0}{(1+(m_g/r_b)^2)^{\frac{3}{2}\beta}} \ ,
\end{eqnarray}
where $T_0$ is the temperature amplitude, $g_c$ is the scaling ratio, $a_c$ is the power index determining the falloff or increase of the temperature at the center, $a_T$ determines the falloff at the large radius, $r_c$ is the core radius and $r_T$ is the temperature scale radius. 
  The conventional beta model, Eqn.~\ref{bmodel}, is used to fit our density points. 
  We define $m_g(R,Z) \equiv \Pi(t_0,\epsilon_2,\epsilon_3,x,y,z)$, as the elliptical coordinate, where $\epsilon_2$ and $\epsilon_3$ are the ellipticities of the gas, where $R$ is the radial coordinate in the plane of sky, $R = \sqrt{y^2 + z^2}$, and $Z$ is the coordinate in the LoS, $Z = z$ (note: $\epsilon_2$ and $\epsilon_3$ are functions of the radius). 
  Because the results of the X-ray analysis are nearly insensitive to the ellipticities of the gas distribution we consider here, we show the best-fitted parameters of the temperature and density functions for the spherical case only (see Table.~\ref{TNEparameter}, and we fixed $a_c$ = 2.0).
  The curves based on the best-fitted parameters assuming spherical symmetry for the temperature, electron number density profiles are shown in Fig.~\ref{TNe}.
\begin{table}
\centering
\caption{Best Fitted Temperature and Density Parameters\label{TNEparameter}}
\begin{tabular}{|c|c|c|c|c|c|} \hline
$T_0$ &$g_{c}$ &$r_c$ &$a_c$ &$r_T$ &$a_T$ \\ 
(keV) &        &(Mpc) &      &(Mpc) &      \\ \hline
$11.6$    &$-0.65$        &$0.107$     &fixed to &$0.478$ &$1.711$ \\ 
$\pm 1.37$  &$\pm 0.0352$ &$\pm 0.009$ &$2.0$    &$\pm 0.079$ &$\pm 0.31$ \\ \hline
\end{tabular}
\begin{tabular}{|c|c|c|} \hline
$n_0$               &$r_b$           &$\beta$         \\ 
$(10^{-2} \#/cm^3)$ &($10^{-2}$ Mpc) &                \\ \hline
$9.24 \pm 0.43$        &$3.35 \pm 0.145$     &$0.61\pm0.0074$ \\ \hline
\end{tabular}
\end{table}

\begin{figure}
\centering
\includegraphics[angle=-90,width=0.48\textwidth]{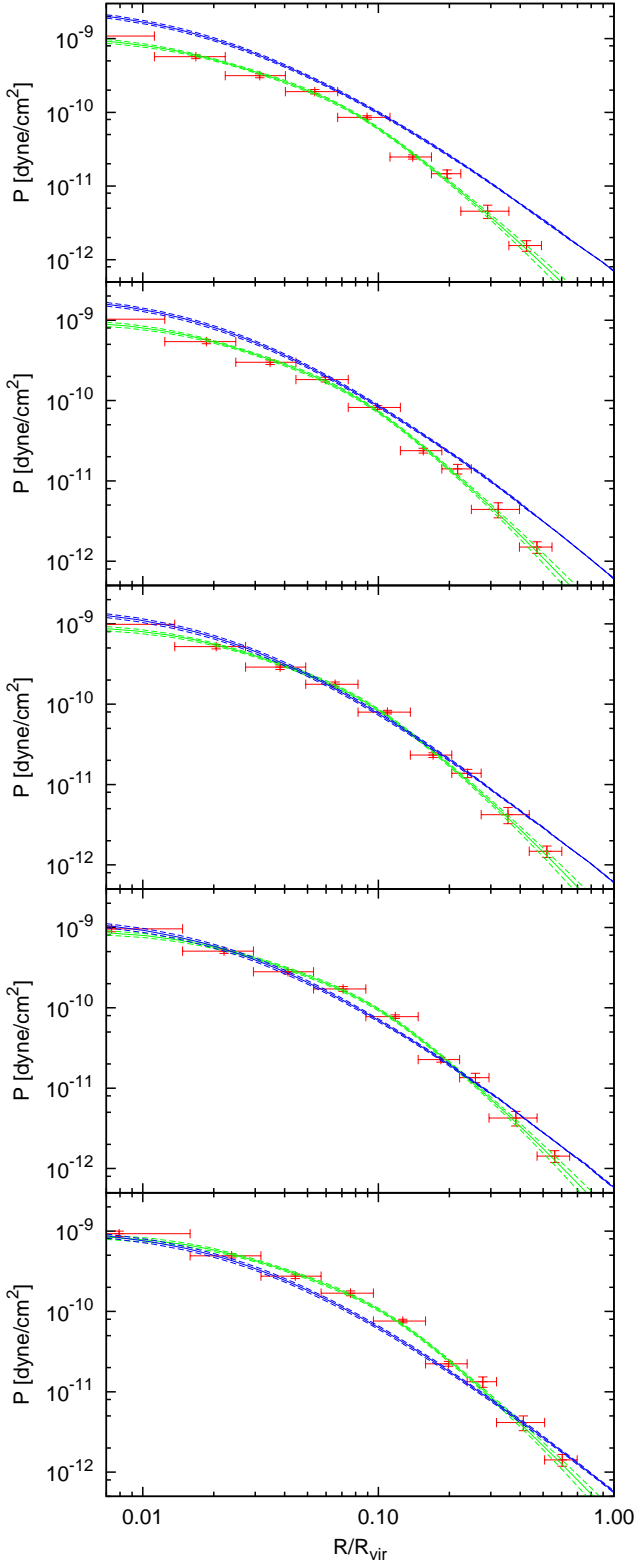}
\caption{
Pressure profiles as a function of projected radius (in the plane of the sky) for different axis ratios \qDM of the dark matter mass distribution (from top to bottom: \qDM = 1.0, 1.25, 1.5, 1.75 and 2.0).
We show the observed pressure distribution (red points with error bars), and 
the best fitted profiles of the thermal and equilibrium pressures (green and blue solid and dashed lines).
\label{Pr}
}
\end{figure}

  We use the best-fitted temperature and electron density profiles to derive the distribution of the thermal pressure using the ideal gas law, $\Pth=n k_B T_g$, where \Pth is thermal gas pressure, $T_g$ is the temperature of the gas assumed to be equal to the electron temperature and $n$ is the total gas number density, the sum of the electron and ion densities, $n=n_e+n_i$ (we assume\footnote{see for example M10} $n_e/n_i=1.1737$). 
   The thermal pressure profiles for each mass model with different ellipticities are shown on Fig.~\ref{Pr}.

%
%

\begin{table}
\centering
\caption{Best-fitted Mass Model Parameters\label{NFWparameter}}
\begin{tabular}{|c|c|c|c|c|c|} \hline
\qDM    &$C_{200}$     &$M_{200}$  
        &\ECvir        &\EMvir       &\ERvir \\ 
            &              &$(10^{+15}\Msun)$ 
            &              &$(10^{+15}\Msun)$ &$(\mbox{Mpc})$ \\ \hline
1.0         &$11.73$ &$0.772$  
            &$14.11$ &$0.847$        &$2.05$  \\ 
            &$\pm0.55$ &${}^{+0.047}_{-0.042}$  
            &$\pm0.65$ &$\pm0.0505$  &        \\ \hline
1.25        &- &-  
            &$12.75$ &$0.77$         &$1.85$  \\ 
            &  &  
            &$\pm0.52$ &$\pm0.050$   &        \\ \hline  
1.5         &- &-  
            &$11.60$ &$0.70$         &$1.68$  \\ 
            &  &  
            &$\pm0.50$ &$\pm0.052$   &        \\ \hline 
1.75        &- &-  
            &$10.744$  &$0.647$      &$1.556$ \\ 
            &  &  
            &$\pm0.477$&$\pm0.055$   &        \\ \hline 
2.0         &- &-  
            &$9.92$     &$0.594$      &$1.446$ \\ 
            &  &  
            &$\pm0.406$ &$\pm0.046$   &        \\ \hline           
\end{tabular}
\tablecomments{
  The mass model parameters.
  First column: the axis ratios \qDM (axis aligned LoS over perpendicular to it) we choose to study. 
  $\qDM=1.0$ implies the spherical symmetry.
  Second and Third column: the NFW model parameters $M_{200}$ and $C_{200}$ in G5. 
  Fourth, Fifth and Sixth column: converted spherical or elliptical NFW model parameters, \ECvir, \EMvir and \ERvir, in term of virialized states.
}
\end{table}

%
%

\section{Equilibrium Pressure}

  In this section we describe our method to derive the equilibrium pressure, \Peq, using published results from a combined analysis of weak and strong lensing observations. 
  The non-spherical geometry of the galaxy clusters has been extensively studied by simulations or observations, and it has been suggested that the gas distribution is most likely different to the total mass distribution assuming hydrostatic equilibrium \citep{Buote1998,JingSuto2002,LeeSuto2003,Oguri2004,Lau2011}. 
  Therefore we derive the gravitational potential using spherical and non-spherical geometry assuming that the total mass density distribution has fixed ellipticities, $e_2$ and $e_3$ (which are not the same as the gas ellipticities: $\epsilon_2=\epsilon_2(e_2,e_3,x,y,z)$ and $\epsilon_3=\epsilon_3(e_2,e_3,x,y,z)$), and then obtain the gas density ellipticities and the gravitational force from the total potential. 
  Finally the equilibrium pressure is derived assuming the hydrostatic equilibrium equation.

%
%

\subsection{Spherical and Non-spherical Deprojection}

  A gravitational lensing analysis provides the projected (2D) distribution of total mass surface density. 
  In order to determine the 3D distribution of the cluster, we need to deproject the 2D mass surface density.

  For both cases, spherical and non-spherical symmetry, we use the more general elliptical NFW model parametrized by $m_{N}(R,Z) \equiv \Pi(t_0,e_2,e_3,x,y,z)$:
\begin{equation}
\rho_{N}(m_{N}(R,Z))=\frac{\rho_{0}}{(\frac{m_{N}}{r_s})(1+\frac{m_{N}}{r_s})^2} \ ,
\end{equation}
where $\rho_0$ is the central density and $r_s$ is the scale radius, which is reduced to the spherical case for $e_2=e_3=0$.

  We use the published best-fitted NFW parameters ($M_{200}$ and $C_{200}$) from G5, and simulate the mass surface density maps based on their best-fitted parameters. 
  Under our assumption of prolate models with the non-degenerated axis aligned in the LoS (i.e. $e_2=e_3$), the surface mass density $\Sigma$ is circularly symmetric in the plane of sky and can be expressed as the function of $R$ as the following:
\begin{eqnarray}
\label{sigma}
\Sigma(R)     &\equiv    &\int^{+\infty}_{-\infty}{\rho_{N}(m_{N}(R,Z)) dZ} \nonumber \\ 
              &=         &2\int^{+\infty}_{m=R}{\rho_{N}(m_{N}) \frac{\qDM \, m_{N}}{\sqrt{m_{N}^2-R^2}}dm_{N}} \ ,
\end{eqnarray}
\qDM is the axis ratio (axis in the LoS over perpendicular to it) defined by $\qDM \equiv 1/\sqrt{1-e^2_2} = 1/\sqrt{1-e^2_3}$ for our prolate models.
  For $\qDM=1.0 \, (\mbox{or} \, e_2=e_3=0)$, Eqn.~\ref{sigma} is reduced to the spherical geometry.

  For our prolate models, we estimate the elliptical NFW model parameters - the elliptical virial mass \EMvir and the elliptical concentration parameter \ECvir - by fitting the models to the mass surface density map. 
  We found that there is no significant effect on \EMvir and \ECvir from how we choose the simulated data points for fitting, thus we generated 15 simulated $\Sigma$ data points distributed uniformly from 0.01 to 3.0 Mpc for our elliptical fitting.
  The errors for the best-fitted parameters, \EMvir and \ECvir, were generated from 1000 Monte-Carlo realizations.
  We show our best-fitted mass model parameters in Table.~\ref{NFWparameter}.
  Once \EMvir and \ECvir are obtained, we define the elliptical virial radius \ERvir based on the average density enclosed within the prolate ellipsoid with projected radius \ERvir, as 
\begin{equation}
  \EMvir \equiv \frac{4\pi}{3} \, (\ERvir)^{3} \, \qDM \, \rho_{m}(z)\Delta(z), 
\end{equation}
where $\rho_{m}(z)$ is the mean matter density and $\Delta(z)$ is the virial overdensity at redshift $z$.

\subsection{Derivation of Equilibrium Pressure}

  We derive the equilibrium pressure \Peq using the equation of the strict hydrostatic equilibrium: 
\begin{equation}
\label{HE_vec}
\vec{\nabla} \Peq = -\rho_{g}\vec{\nabla} \Phi,
\end{equation}
where $\rho_{g}$ is the gas mass density and $\Phi$ is the total gravitational potential.
  We derive the gravitational force from the total gravitational potential $\Phi$. 
  For a given NFW mass distribution $\rho_N$ and a set of the total mass ellipticities $e_2$ and $e_3$, $\Phi$ can be written as the following \citep{BinneyTremaine2008}:
\begin{eqnarray}
\label{Pot}
\Phi(e_2,e_3,x,y,z) &= &-\pi \, G \, \sqrt{1-e_2^2} \sqrt{1-e_3^2} \times \\
& &\int_{t=0}^{\infty}{dt \left[ \frac{\psi(\infty) - \psi(\Pi^2)}{\sqrt{t+1}\sqrt{t+1-e_2^2}\sqrt{t+1-e_3^2}} \right] }  \nonumber \\
\psi(\Pi^2) &\equiv &\int_{0}^{\Pi^2}{dm_N^2 \, \rho_N(m_N^2)}. \nonumber
\end{eqnarray}

  For our case of prolate models, Eqn.~\ref{HE_vec} can be parameterized using $R$, and reduces to 
\begin{equation}
\label{hydro} 
\frac{1}{\rho_{g}}\frac{d\Peq}{dR} = g(R),
\end{equation}
where $\rho_{g}$ is the gas mass density.
  Using Eqn.~\ref{Pot}, $g(R)$ can be written as the following:
\begin{eqnarray}
\label{g}
g(R) &= &-\frac{4 \pi G }{ \qDM } \int_{m_N=0}^{\qDM \, R}{\frac{m_N^2 \rho_{N}(m_N) dm_N}{\left[ R^2+m_N^2 (1-\frac{1}{\qDM^2}) \right]}} \nonumber \\ 
     &= &-4\pi G \rho_0 r_s \qDM \times \nonumber \\
     &  &\int^{\frac{\qDM \,R }{r_s}}_{0}{\frac{x dx}{(1+x)^2}\frac{1}{(\frac{\qDM \, R}{r_s})^2 + x^2(\qDM^2-1)}} \ ,
\end{eqnarray}
where $x \equiv m_N/r_s$ and $G$ is Newton's constant. 
  For \qDM = 1, Eqn.~\ref{g} reduces to the expression for spherical geometry.

  Using the best-fitted NFW model parameters and the gas density profiles derived from X-ray analysis (assuming the mean molecular weight $\mu=0.61879$, as in M10), the equilibrium pressure, \Peq, profiles can be derived numerically by integrating Eqn.~\ref{hydro}. 
  Nevertheless, Eqn.~\ref{hydro} is a first-order differential equation thus one boundary condition is needed for a solution. 
  Unfortunately, the pressure is not known neither at the cluster center, nor at the virial radius (where it is not decreasing to zero due to the accretion shock).

%
%
\begin{figure}
\centering
\includegraphics[angle=-90,width=0.49\textwidth]{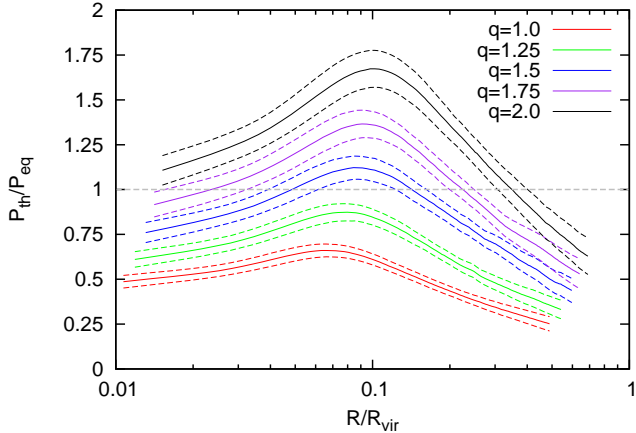}
\caption{
Pressure ratio profiles $\Pth/\Peq$ for MS~2137 assuming axis ratios of \qDM = 1.0, 1.25, 1.5, 1.75 and 2.0 (red, green, blue, purple and black solid and dashed lines).
The horizontal grey dashed line indicates hydrostatic equilibrium. 
\label{Rpr}
}
\end{figure}

  Analyzing AMR and SPH simulations, Molnar et al. have found that the accretion shocks in clusters are located at 0.9--1.3 \Rvir, where the pressures in different clusters are a few times $10^{-13}$ dyne/cm$^2$ (\citealt{Molnar2009}; 2010).
  Currently, we cannot derive the pressure at \Rvir due to the very weak X-ray signal from large radii, therefore we have no constraints on any boundary conditions from observations. 
  Fortunately, the pressure at \Rvir, $\Peq(R=\Rvir)$, contributes less than 0.1\% to the central pressure, therefore the choice of its value has no significant effect on the pressure in the inner part of the cluster.
  However, deriving the pressure distribution based on different boundary conditions (different $\Peq(R=\Rvir)$), we noticed that the slope of the pressure profile changes only close to \Rvir, and it is about about 0.8 near \Rvir = 0.7. 
  Therefore assuming the same slope for radii close to \Rvir provides a smooth pressure distribution, which we would expect out to \Rvir (see M10 for the example). 
  The equilibrium pressure profiles we obtained based on the assumption above are shown in Fig.~\ref{Pr}.

%
%
\section{Discussion}

  \subsection{Testing Hydrostatic Equilibrium}
  We test the assumption of strict hydrostatic equilibrium in MS~2137 assuming spherical and prolate models by analyzing the pressure ratio profiles $\Rpr \equiv \Pth/\Peq$. 
  We use five different axis ratios for the dark matter distribution (the semi-axis along LoS over the perpendicular to it): \qDM = 1.0 (spherical case), 1.25, 1.5, 1.75, and 2.0.
  \Rpr = 1.0 implies strict hydrostatic equilibrium, while \Rpr $<$ 1 implies that non-thermal pressure, $\Pnth \equiv \Peq - \Pth$, is required for the cluster to be in equilibrium.
  \Rpr $>$ 1 would imply that the thermal pressure is larger than the pressure required for equilibrium, therefore the cluster is expanding, which is not acceptable for relaxed clusters (there would be no physical reason for the cluster to expand).
  We show the \Rpr profiles as a function of projected radius $R$ in Fig.~\ref{Rpr}. 
  All ratio profiles are plotted only within the radial range where both X-ray and lensing observations were available (no extrapolations were used).

%
%

\begin{figure}
\centering
\includegraphics[angle=-90,width=0.49\textwidth]{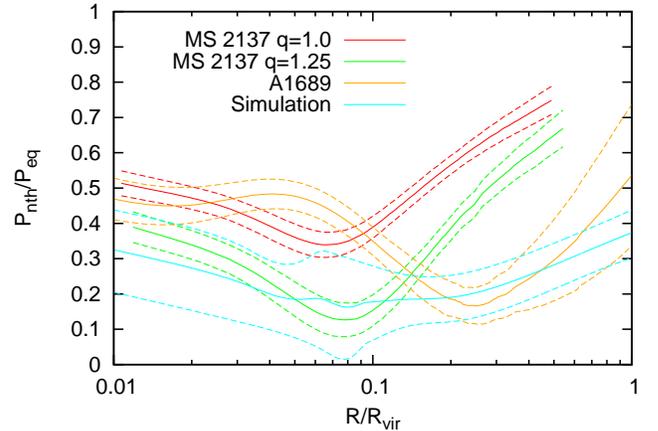}
\caption{
Non-thermal pressure ratio profiles, $\Rnpr \equiv 1-\Pth/\Peq = \Pnth/\Peq$, for MS~2137 with different axis ratios (the assumed axis ratios and color codes are the same as in Fig.~\ref{Rpr}. 
We also show \Rnpr profiles of A1689 and two simulated relaxed clusters studied in M10 under the assumption of spherical geometry  (orange and cyan solid and dashed lines).
\label{nRpr}
}
\end{figure}

  Assuming a spherical model for MS~2137, the thermal pressure contribution is about 50\% $\pm$ 5\% within 0.03 \Rvir, and it increases to a maximum of about 60\% at 0.08 \Rvir, and decreases again to about 25\%--40\% at 0.2--0.5 \Rvir. 
  In MS~2137, overall, the thermal pressure contribution is almost half of the equilibrium pressure at all radii, which is consistent with the result of G5. 
  As a result, the mass determined from X-ray analysis by assuming spherical symmetry and hydrostatic equilibrium is systematically less than the mass based on the lensing analysis by 40\% at all radii. 
  As it was shown by G5, a possible reason for this discrepancy is non-spherical geometry. 
  We find that, assuming prolate models with different axis ratios, \qDM (principal axis aligned with the LoS), the non-thermal pressure contribution necessary for equilibrium is reduced, and thus the two mass estimates would be closer.
  From Fig.~\ref{Rpr}, \qDM $\simeq$ 1.25 can result less non-thermal contribution at all radii and be closest to the hydrostatic equilibrium at about 0.1 \Rvir, implying that MS~2137 could possibly be far from the spherical symmetry.
  Assuming \qDM = 1.75, we can resolve the discrepancy within 0.03 \Rvir for MS~2137 but non-physically overestimating the \Pth at 0.07--0.2 \Rvir.
  Setting \qDM = 2.0 for MS~2137 overestimates the \Pth significantly at almost the whole radii.
  However, for the large radii beyond 0.5 \Rvir, MS~2137 is deviated significantly from the hydrostatic equilibrium among \qDM ranging from 1.0 to 2.0.

  We show the non-thermal pressure ratio profiles, $\Rnpr \equiv 1-\Rpr$ (or $\Pnth \equiv \Peq-\Pth$), for MS~2137 in Figure~(\ref{nRpr}).  
  For comparison, in this figure, we also show non-thermal pressure ratio profiles of Abell~1689 and simulated clusters using results from M10 (where we assumed spherical symmetry).
  The \Rnpr values for MS~2137, assuming spherical symmetry, are systematically higher than the those derived from simulations by 2 $\sigma$, providing further evidence that spherical symmetry for MS~2137 is not a good assumption. 
  Assuming \qDM = 1.25--1.5, \Rnpr of MS~2137 we find a good match with the simulations within 0.1 \Rvir, but they deviate at the large radii (beyond 0.15 \Rvir), suggesting that the non-thermal pressure components could possibly contribute significantly at the large radii for MS~2137.
  Although the assumption of hydrostatic equilibrium in MS~2137 seems to be better if we adopt \qDM = 1.25--1.5 (statistic value is 11740.0 with 10632 degrees of freedom, reduced $\chi^2=$ 1.10 for \qDM = 1.25), we still find a significant non-thermal contribution of about 10\%--40\% in the core region. 
  Our results suggest that, even if we assume elliptical models, strict hydrostatic equilibrium is not valid in the core region of galaxy clusters, and therefore the contribution from non-thermal pressure has to be taken into account when accurate cluster mass determination from X-ray observations is preformed.

  Interestingly, the \Rnpr profile for A1689 seems to be qualitatively different from the profile derived for MS~2137 and from those derived from simulations.
  While the \Rnpr profiles for MS~2137 and those from simulations monotonically decrease to their minimum at about 0.1 \Rvir, and increase at large radii, \Rnpr for A1689 is roughly constant in the center region (within about 0.06 \Rvir), decreases to its minimum at about 0.25 \Rvir, and increases monotonically for large radii.
  The minimum non-thermal pressure contribution in Abell~1689 located at around 0.25 \Rvir is about twice as far as those in MS~2137 and simulated relaxed clusters.
  Our analysis of pressure profile ratios suggest that Abell~1689 is not a typical relaxed cluster.
  Possible explanations for the peculiar pressure ratio distribution in Abell~1689 can be due to substructure/ the lack of dynamical equilibrium in the core of Abell~1689, and/or feedback from heating, which could push the most relaxed regions outward in the cluster.
  Clearly, more detailed studies are needed to understand the structure of Abell~1689.

\subsection{Sources of Systematic Bias}

  Since we are comparing the equilibrium pressure, which depends directly on the lensing measurement, to the thermal pressure which depends directly on the X-ray data, our results are sensitive to the calibration of the two measurements. 
  In practice, we find that most of our systematic errors are coming form the lensing mass measurement.
  Several sources of systematic bias in lensing measurements have been identified, such as 
 the inherent shapes of the background galaxies, signal dilution due to cluster galaxies, the low signal-to-noise ratio, substructure, contamination by mass concentrations along the LoS, asphericity, the assumed mass models \citep{Rozo2010,Oguri2010,Oguri2011PhRvD,Schmidt2011,Gruen2011,Becker2011,Bahe2012}.
  The accumulated systematic errors due to the uncertainty of the background galaxies and the mass concentration outside $R_{200}$ is about 10\% \citep{Bahe2012}, resulting the same order bias to the pressure ratios.

  Assuming spherical clusters, the dominant systematic bias is originated form triaxiality, which may result an error in the mass measurement more than 40 -- 50\% \citep{Corless2008,Hamana2012,Sereno2012_b,Bahe2012}.
  In order to avoid this main source of uncertainty, we determine pressure profiles using ellipsoidal models with different values for the ellipticity, \qDM, and to have a conservative estimate on the non-thermal pressure support, we accept the value of \qDM, which provides the least amount of non-thermal pressure support without violating our assumption of equilibrium.
  This method is not sensitive to systematic bias in the total mass determination due to substructure or mass concentrations in the LoS.

  On the other hand, the X-ray analysis is not affected significantly by the assumed geometry of the cluster.
  However, clumping of the gas in the outskirts would enhance the X-ray emission resulting in overestimate in the temperature and the density of the gas \citep{Mathiesen1999,Nagai2011,Eckert2012}.
  Shocks would result an excess in the X-ray emission and an overestimate of the thermal pressure, resulting a lower non-thermal contribution. 
  Of course, our assumption of equilibrium would be wrong for merging clusters.
  We choose to use MS~2137 specifically to avoid most of these systematic biases. 
  MS~2137 shows no sign of a recent merger event, no significant substructure, and it looks nearly circularly symmetric in projection. 
  We avoid systematic bias due to extrapolation of our parametrized models to radii we do not have data by studying the cluster in the radial region observations cover. 
  Therefore we do not expect significant bias in our results.

\subsection{Sources of Non-Thermal Pressure Support}

  Several physical mechanisms have been proposed for non-thermal pressure support in the last few decades. 
 The most likely sources are magnetic, cosmic ray pressure and pressure due to turbulent gas motion.
  Magnetic fields in the clusters could grow from the hierarchical formations or could be injected by galactic winds, then possibly further undergo galactic dynamo processes \citep{Goncalves1999,Dolag2000,Govoni2004,Churazov2008}.

  Magnetic fields in the core of several clusters of galaxies have been estimated using Faraday rotation measurements, and found to be in the range of 10--30 (2--8)$\mu G$ for (non-)cool core clusters (i.e., \citealt{Taylor1993,Feretti1999,Taylor2002}).
  It has been suggested that the magnetic field strength is proportional to the density \citep{Dolag1999,Colafrancesco2007,Ando2008}.
  Using simulations, \cite{Dolag2000} have found that the averaged magnetic pressure is about 5\% of the thermal pressure in relaxed clusters, implying that other sources of pressure are needed to explain the discrepancy between the X-ray and lensing mass estimates.

  Shocks generated by structure formation, active galactic nuclei (AGNs) and supernova feedback may accelerate particles to very high energies.
  It has been suggested that cosmic rays generated by these processes could also provide a source for non-thermal pressure support \citep{Ensslin1997,Sarazin2004,Guo2008}.
  Constraints on the cosmic ray pressure contributions can be placed by studying the gamma rays emission generated by neutral pions produced by collisions between cosmic ray and ICG particles \citep{Ando2008}. 
  Based on numerical simulations and observations, the contribution from cosmic rays pressure was estimated to be less than 10--30\% of the thermal pressure in the very central region (within about few tens kpc) of clusters \citep{Ensslin1997,Miniati2001,Reimer2003,Pfrommer2004,Pfrommer2007_1,Pfrommer2008_2,Pfrommer2008_3,Sijacki2008}.

  The most widely accepted candidate for the non-thermal pressure support is turbulent gas motion. 
  Cosmological simulations show that even relaxed clusters contain 100 kpc scale subsonic flows and turbulence generated by structure formation \citep{Evrard1990,Norman_Bryan1998,Nagai2007,Fang2009,Lau_Kravtsov_Nagai2009,Vazza2011}. 
  At the outskirts of clusters subsonic flows are most likely generated by previous major mergers, and additional turbulence can be generated by supersonic motion of cluster galaxies. 
  Accretion via filaments delivers gas and galaxies to the core region of clusters providing power for turbulence in the core. 
  Non-gravitational processes such as jets and bubbles from AGNs may also contribute to turbulence in the core region \citep{Churazov2002}.
  Based on simulated clusters, M10 showed that even without additional non-gravitational physics the non-thermal pressure support in the core of relaxed clusters can be about 30\%, reaches a minimum in the core region (at about 0.1 \Rvir), and increases to about 35\% at the virial radius. 
  M10 also showed, that assuming spherical models, there is a significant contribution, about 35\%, from non-thermal pressure at the central region of A1689.
  In this paper we have shown that, even if we use elliptical cluster models, 
the strict hydrostatic equilibrium is not valid in MS~2137, we find a significant, 30\%--40\% contribution to non-thermal pressure in the core region. 
  Our result indicate that the significant non-thermal pressure support in the core region of clusters is not a consequence of the assumed spherical shape of the clusters, but a general feature.

  A detailed study of the sources of different non-thermal pressure contribution in MS~2137 would require more detailed observations in different wavelengths, and therefore it is out of the scope of our paper. 
  In principle, the different non-thermal pressure contributions could be estimated using parametrized functions, but the results would strongly depend on the parametrization \citep{Lagana2010}. 
  The quality of MS~2137 observations available to us would not allow a meaningful separation of the non-thermal pressure contributions, therefore we have presented our results in terms of thermal and non-thermal pressure support.

%
%
\section{Conclusion}

  We have derived thermal and non-thermal pressure ratio profiles for cluster MS~2137 from X-ray and gravitational lensing observations assuming spherical and elliptical models. 
  Since MS~2137 show nearly circular projected morphology, with a good approximation, in our analysis, we could assume elliptical (prolate) models with the non-degenerate axis aligned in the LoS.

  Analyzing the pressure ratio profiles, we have found, in accord with previous studies based on different arguments, that a prolate model with \qDM = 1.25--1.5 for MS 2137 provides an adequate description for the distribution of the ICG. 
  We have found that the contribution from non-thermal pressure varies with radius.
  Based on our analysis of the pressure ratios (\Rpr and \Rnpr), spherical symmetry does not seem to be a good assumption for MS~2137. 
  Assuming \qDM = 1.25--1.5, \Rnpr reaches its minimum value of 0\%--20\% at about 0.08 \Rvir, and the non-thermal component contributes about 10\%--40\% and 20\%--60\%  at small (within 0.03 \Rvir) and large (0.2--0.5 \Rvir) radii, respectively.
  We conclude that MS~2137 is not in hydrostatic equilibrium in these two regions even assuming non-spherical symmetry.
  Our results from comparing the shapes of pressure ratio profiles, \Rnpr, for MS~2137, A1689, and those derived from relaxed clusters drawn from cosmological simulations, suggest that A1689 is not a typical relaxed cluster.

  Our results based on X-ray and gravitational lensing observations, and numerical simulations of relaxed massive galaxy clusters provide further evidence that there is a significant contribution from non-thermal pressure in the core region of even relaxed clusters, i.e., the assumption of hydrostatic equilibrium is not valid in this region, independently of the assumed shape of the cluster (spherical or elliptical).

\acknowledgements
INC thanks K. Umetsu for discussions regarding the lensing analysis, and H.-H. Chan for advice on deprojection techniques. 
We thank the referee for comments and suggestions which helped us to improve our paper.

\clearpage

%
%
\begin{appendix}

  We project our gas models onto the plane of sky assuming that the gas is in the equilibrium with the gravitational potential generated by our assumed total mass distribution.
  In our mass model, the dark matter density distribution is described by a series of mass shells with constant axis ratios, \qDM (principal axis along LoS over the minor axis, which is perpendicular to LoS).
  The equipotential surfaces generated from each mass shells coincides with a series of the confocal ellipsoidal surfaces, and the axis ratio for these equipotential surfaces is getting closer to 1 (the shapes are becoming rounder as the radius increases).
  We derive the total potential by summing up the potential generated from each mass shell.
  As a consistency test, we checked how much the total potential changes once we add the gas to the system, and found that the change is insignificant, less than 0.1\%. 
  We concluded that the geometry of the equipotential surfaces is dominated by the dark matter halo, so we can ignore the effect of the gas.
  We assume the gas distribution is identical to shape of the total potential by the X-ray Shape Theorem \citep{Buote1998}.

   Before projection, we define our coordinate system which is used throughout this paper. 
   We define the confocal ellipsoidal coordinates, $\Pi(t,e_2,e_3,x,y,z)$, by the following:
\begin{eqnarray}
\Pi^2(t,e_2,e_3,x,y,z)   &=   &\frac{x^2}{t+1}+\frac{y^2}{t+1-e_2^2}+\frac{z^2}{t+1-e_3^2}  \\
e_2^2                    &=   &1-(a_2/a_1)^2 \nonumber \\
e_3^2                    &=   &1-(a_3/a_1)^2 \nonumber
\ ,
\end{eqnarray}
where $a_1$,$a_2$ and $a_3$ are the three semi-axes and $a_3 \leq a_2 \leq a_1$ (i.e. $0 \leq e_2 \leq e_3 < 1$). $x$, $y$ and $y$ are the coordinates of three semi-axes, $t$ is the parameter representing the family members of the same series of the confocal surfaces. 
  Given a fixed $e_2$ and $e_3$, the shape of $\Pi$ surface becomes rounder as $t$ increases. 
  For $e_2=e_3=0$, $\Pi$ is reduced to the spherical distribution. 
  For $t$ = constant, 
  $\Pi$ describes a conventional ellipsoidal distribution.

  The emissivity in clusters is dominated by the thermal bremsstrahlung radiation and is equal to the product of the square of the gas number density, $n_e^{2}$, and the cooling function, $\Lambda$. 
  For convenience, we label the inner most to the outer most annuli, where we extract the spectra from, by 1 to N, consecutively. 
  Assuming the gas is homogeneous within each shell, the expected luminosity we observed for the $k$th annulus in the plane of sky, $L_{k}$, can be expressed as:   
\begin{equation} 
\label{Luminosity}
    L_{k} = \Sigma_{i=k}^{N} {n_{e}}_{i}^2 \Lambda_{i} V_{k,i} \ , 
\end{equation}
where $i$ represents the quantity of the $i$th shell, and $V_{i,j}$ is the volume of the $j$th shell occupying the cylinder having the non-degenerated axis aligned LoS and the base of the $i$th annulus.
  By introducing the volume ratio $R_{k,i} \equiv V_{k,i}/V_{i,i}$, the Eqn.~\ref{Luminosity} can be reduced to the following form:
\begin{equation} 
\label{rdL}
L_{k} = \Sigma_{i=k}^{N} ( {n_{e}}_{i}^2 \Lambda_{i} V_{i,i} ) R_{k,i}
\end{equation}

   Eqn.~\ref{rdL} can be viewed as the basic ideal of the spectral fitting. 
   $V_{i,j}$ as well as $R_{i,j}$ ($i,j=1,2,\ldots,N$) can be computed directly from the gas distribution.
   We derive our gas models in the 3D space and project them onto the plane of sky for a given set of volume ratios $R_{k,i}$, which connect the normalization parameters in the \textit{MEKAL} models of the different shells. 
   We fit the models to the spectra simultaneously to get the temperature, the abundance and the normalization parameter of each shell.
   
\end{appendix}


\bibliographystyle{apj}

\end{document}